\documentclass[12pt,preprint]{aastex}

\shorttitle{Point Process Algorithm}
\shortauthors{Marsh et al.}

\begin{document}

\title{Point Process Algorithm: A New Bayesian Approach for Planet
Signal Extraction with the Terrestrial Planet Finder Interferometer}

\author{Kenneth A. Marsh\altaffilmark{1}, Thangasamy Velusamy\altaffilmark{1}, 
Brent Ware\altaffilmark{1}}

\altaffiltext{1}{Jet Propulsion Laboratory, 4800 Oak Grove Drive, Pasadena,
CA 91109; Kenneth.A.Marsh@jpl.nasa.gov, 
Thangasamy.Velusamy@jpl.nasa.gov, Brent.Ware@jpl.nasa.gov}

%\pagebreak

\begin{abstract}
The capability of the Terrestrial Planet Finder Interferometer
(TPF-I) for planetary signal extraction, including both
detection and spectral characterization, can be optimized by
taking proper account of instrumental characteristics and
astrophysical prior information. We have developed the Point
Process Algorithm (PPA), a Bayesian technique for extracting
planetary signals using the sine-chopped outputs of a dual
nulling interferometer.  It is so-called because it represents the
system being observed as a set of points in a suitably-defined
state space, thus providing a natural way of incorporating our
prior knowledge of the compact nature of the targets of interest.
It can also incorporate the spatial covariance of the exozodi as
prior information which could help mitigate against false
detections. Data at multiple wavelengths are used simultaneously,
taking into account possible spectral variations of the planetary
signals.  Input parameters include the RMS measurement noise
and the {\em a priori\/} probability of the presence of a planet.
The output can be represented as an image of the intensity
distribution on the sky, optimized for the detection of point
sources. Previous approaches by others to the problem of planet
detection for TPF-I have relied on the potentially non-robust
identification of peaks in a ``dirty" image, usually a correlation
map. Tests with synthetic data suggest that the PPA provides
greater sensitivity to faint sources than does the standard
approach (correlation map $+$ CLEAN), and will be a useful tool
for optimizing the design of TPF-I. 
\end{abstract}


\section{Introduction}

\begin{figure}
 \includegraphics[scale=0.65, angle=0]{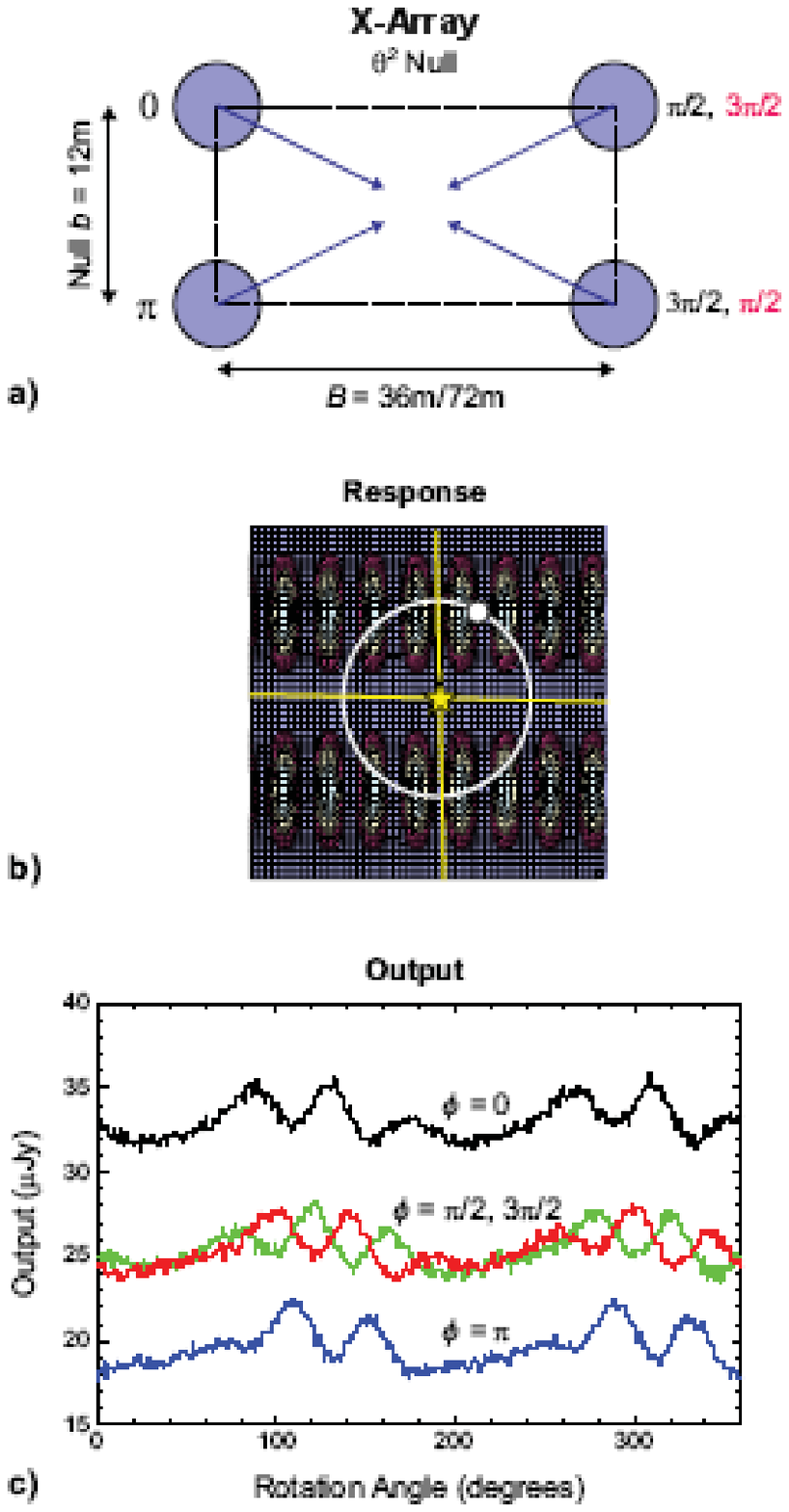}
  \caption{Schematic of Dual-Bracewell Nulling Interferometer array in the 
X-configuration.
(a) Assumed geometry: Two 12 m nulling pairs are separated by 36 m or 72 m.
Within each nulling pair, the signals are combined with a relative phase shift 
of $\pi$, resulting in destructive interference at the location of the star.  
The signals from the two nulling pairs are then combined coherently after
introducing relative phase shifts, alternately, of $\pi/2$ and $3\pi/2$;
the difference between these two signals then constitutes the ``sine-chop"
signal.  In practice, the required pair of sine-chop differencing signals is 
obtained by combining the beams from all four elements simultaneously in a 
central beam combiner, after introducing a set of
phase shifts, to the individual elements, of either $\{0,\pi,\pi/2,3\pi/2\}$ or
$\{0,\pi,3\pi/2,\pi/2\}$.
(b) Angular response on the sky when combining the two nulled signals
with $\pi$ phase shift; the star (nulled out) is at the center,
and the planet position is indicated by the white disk.
A modulated signal is produced as the interferometer system rotates and the
planet passes through peaks and valleys of the response.
(c) Example of modulated signals at 12 $\mu$m wavelength simulating 
an observation of a planet 3 Earth radii in size, 1 AU from a G2 star at 
10 pc, using an X-array with $B=36$ m and $b=9$ m.} \label{fig1}
\end{figure}

A key component of the proposed Terrestrial Planet Finder mission
is a nulling interferometer (TPF-I) for the wavelength range 7--15
$\mu$m,  currently envisaged as a free-flying four-element dual-Bracewell 
array in the ``X" configuration as shown in Figure \ref{fig1}.  The signals 
from the two elements in a nulling pair are combined with $\pi$ phase shift, 
resulting in destructive interference for a source on the optical axis, i.e., 
the axis corresponding to equal geometrical path lengths to the two nulling
elements.  By placing this null at the location of the star, the contrast
between planets and star is dramatically increased, thereby overcoming
one of the chief obstacles to planet detection \citep{brace78}.  
The starlight suppression is further improved by using two such nulling pairs 
and combining their outputs coherently (see, for example, 
\citet{beichman_velusamy1999}). As the interferometer system is rotated about 
the line of sight, a modulated signal is then produced, and this signal
contains information on the intensity distribution of sources on the sky.
Its information content can be increased by alternately 
applying phase shifts of $\pi/2$ and $3\pi/2$ and taking the difference of 
the phase-shifted signals. This has the effect of suppressing
instrumental effects such as low-frequency detector noise and thermal
background drifts, and is commonly referred to as ``sine chopping" 
(\citep{wol98,velusamy_beichman2001}). Sine chopping also suppresses any
structure on the sky which is symmetrical about the star, including that
from symmetrical structure in the exo-zodiacal (exozodi) cloud, thus
increasing the contrast for the detection of planetary signals.

Since the interferometric signals from such a small number of baselines 
provide very limited information on spatial Fourier components of the
intensity distribution of sources on the sky, it is necessary to supplement
the observations with {\em a priori\/} information in order to recover
that distribution.  One widely-used procedure for
this purpose is the maximum entropy method (see \citet{nn86} for a review, and
\citet{sut06} for recent developments).  However, such procedures are not
optimal for the present problem because they do not incorporate an important
piece of prior knowledge, namely the pointlike nature of the planetary
sources.

Of the various inversion procedures that do incorporate such information
(see, for example, \citet{angel_woolf1997, vel04}), the one most widely used 
involves making a correlation map (referred to as a ``dirty image") followed
by deconvolution with the CLEAN algorithm \citep{hog74,draper_etal2005}. 
A key step in that procedure is an iterative search for peaks in the
dirty image, which can be non-robust when noise bumps fall on ``side
lobes" of the responses to other sources.

In this paper we propose a Bayesian technique for planet detection which
avoids the noise-vulnerable peak-finding step. It can
process data at many wavelengths simultaneously, and can incorporate
prior information such as the spatial covariance of the exozodi.
Since the planetary signals are represented as
a random set of points in a suitably-defined state space, we refer
to the resulting algorithm as a ``Point Process Algorithm'' (PPA) following
the terminology of \citet{ric92}. 

The PPA, whose mathematical foundation is presented in detail by \citet{ric87},
is a general technique of which the TPF problem represents a specific 
application \citep{vel05}.  It may be regarded as a state estimation technique 
in which the system to be described (for example, an image) consists of the 
superposition of an indefinite number of objects, each of
which is characterized by a set of parameter values. Each object can
thus be represented as a single point in a multidimensional state space of
which each axis corresponds to one of the parameter values.  
In our application, the axes of the state space
are simply the $x$ and $y$ positions and flux density of an individual
point source. Other possible examples are: (1) the representation of an image of
a cluster of galaxies in terms of a 6-dimensional state space in which the
axes represent the $x$ and $y$ location, the flux density, major and minor 
axes and orientation of an individual galaxy, and (2) the representation of
a planetary system using a state space whose axes consist of the orbital
parameters characterizing an individual planet, using a set of measurements
of Doppler shifts or astrometric measurements of the parent star (or both).
In each of these examples, the goal of the
PPA is to estimate the most probable set of points in the state space
given the set of measurements.  

\section{Measurement Model}

The starting point of our approach is a measurement model which
relates a data vector, $\mathbf d$ (whose components are the
complete set of samples of the sine-chop signal at all wavelengths
of observation), to the intensity distribution, $I(x,y)$, on the
sky. The latter is modeled as the superposition of a set of point
sources of unknown number, fluxes and positions, upon an extended
background whose intensity at position $\left(x_{j},y_{j}\right)$
is denoted by $\zeta_{j}$.  The distribution of point sources is
represented as a set of occupation numbers in a 3-dimensional
state space whose axes are flux and $x$, $y$ position. The state
space is divided into a regularly-sampled grid of cells, such that
the cell with coordinates $\left(f_{k},x_{j},y_{j}\right)$ represents the
$k$th possible flux value, $f_{k}$, at the $j$th spatial position,
$\left(x_{j},y_{j}\right)$.  If a point source of that particular
flux density is present at that particular position, then the
occupation number of that cell, $\Gamma_{n}$ (where indices $j$ and $k$
have been mapped onto a single index, $n$), will be equal to 1;
otherwise, it will equal zero. This representation provides a
convenient framework within which to incorporate our prior knowledge of spatial
dilution.

We assume that the source distribution is being observed by a rotating 
dual-Bracewell interferometer consisting of two nulling pairs separated by an 
``imaging baseline," and that the measurements consist of a time series of 
sine-chop signal values as the system rotates with respect to the sky. If the 
$i$th such measurement is made at an orientation corresponding to instantaneous
baseline vectors ${\mathbf b}_i$ and ${\mathbf B}_i$ for the nulling and
imaging baselines respectively, then the instantaneous response to a point
source of unit flux at off-axis sky position $(x_j,y_j)$ 
(denoted by unit vector ${\mathbf s}_j$) is given by:
\begin{equation}
H_i(x_j,y_j) = -4 \sin^2(\pi{\mathbf s}_j\cdot{\mathbf b}_i)
\sin(2\pi{\mathbf s}_j\cdot{\mathbf B}_i)R_i({\mathbf s}_j)
\end{equation}
where $R_i({\mathbf s}_j)$ is the ``primary beam" response, i.e., the response
of an individual detector, and the baselines ${\mathbf b}_i$ and ${\mathbf B}_i$
are in units of wavelengths.

The measurement model
must include a spectral model which relates the planetary fluxes at
the different wavelengths.  There are several possibilities,
one of which is to treat the planets
as black bodies at the local radiation temperature.  If the orbital
inclination and position angle of the orbital tilt axis are known,
the planet temperature, $T_{j} \equiv T\left(x_{j},y_{j}\right)$,
can then be determined at each $\left(x_{j},y_{j}\right)$
from knowledge of the stellar luminosity and radial
distance. Our measurement model can then be written as:
\begin{equation} \label{eqn_D}
d_{i} =
\sum_{j,k}{f_{k}\left[{B\left(\lambda_{i},T_{j}\right)}/{B\left(\lambda_{0},T_{j}\right)}\right]H_{i}\left(x_{j},y_{j}\right)\Gamma_{n}}
+
\sum_{j}{H_{i}\left(x_{j},y_{j}\right)\zeta_{j}\left(\lambda_{i}\right)}
+ \nu_{i}
\end{equation}

\noindent where $\lambda_{i}$ represents the wavelength
at which the $i$th measurement is made, $B\left(\lambda,T\right)$
represents the Planck function, and the $f_{k}$ represent
possible values of planetary flux at some suitably-defined reference
wavelength, $\lambda_{0}$, and $\nu_i$ is the measurement noise. The index,
$n$, is understood to be a function of $j$ and $k$.

It will be convenient to lump both of the last two terms on the
r.h.s. of (\ref{eqn_D}) into a single ``noise'' vector, $\mu$, whose
components are given by:
\begin{equation} \label{eqn_E}
\mu_{i} = \sum_{j}{H_{i}\left(x_{j},y_{j}\right)\zeta_{j}} + \nu_{i}
\end{equation}

If we then define a matrix $\mathbf{A}$ whose components are
\begin{equation} \label{eqn_F}
A_{in} = f_{k}\left[{B\left(\lambda_{i},T_{j}\right)}/
{B\left(\lambda_{0},T_{j}\right)}\right] H_{i}\left(x_{j},y_{j}\right)
\end{equation}

\noindent then we can rewrite (\ref{eqn_D}) in matrix notation as:
\begin{equation} \label{eqn_G}
\mathbf{d} = \mathbf{A}\mathbf{\Gamma} + \mu
\end{equation}

\noindent where $\mathbf{\Gamma}$ is the state vector whose
components are $\Gamma_{n}$.

\section{Statistical Assumptions}

The measurement noise in the TPF signals is distributed as a Poisson process,
dominated by the photon statistics of stellar leakage and the local zodi
background.  Since the expected number of photons per integration period is 
large (typically $\sim10^5$), the statistical distribution accurately 
attains the limit corresponding to a Gaussian random process (GRP).  The 
measurement noise, $\nu_{i}$, on the sine-chop signal can then be regarded as 
a zero-mean GRP for which:
\begin{equation} \label{eqn_H}
E\nu_{i}\nu_{i^{\prime}} =
\left(\sigma_{\nu}\right)_{i}^{2}\delta_{ii^{\prime}}
\end{equation}

\noindent where $E$ is the expectation operator and $\delta$ is the
Kronecker delta.  

For present purposes we regard the exozodi as an error source rather than as
a quantity to be estimated. We model it only to the extent necessary to
mitigate its effects on planet detection. If it is distributed
symmetrically about the star, then
its only effect is to increase $\sigma_\nu$ by a factor corresponding
to the square root of the increase in the overall photon count, since its
coherent contribution cancels out in the sine chop signal.  If the exozodi
is not symmetrical, one approach for dealing with it is to subtract
its estimated contribution from the measurements ahead of time, based on
available data from previous (possibly ground-based) observations
at lower resolution.  The error resulting from this subtraction may be
regarded as an additional zero-mean error contribution to our measurement 
model.  It will, in general, be spatially correlated as a result of the finite 
resolution of the observations and the intrinsic spatial correlation properties
of the exozodi itself. We therefore approximate the residual exozodi,
$\zeta_j(\lambda_i)$, as a GRP with covariance $C_{\zeta}$ defined by:

\begin{equation} \label{eqn_I}
E\zeta_{j}\zeta_{j^{\prime}} = \left(C_{\zeta}\right)_{jj^{\prime}}
\end{equation}

From (\ref{eqn_H}) and (\ref{eqn_I}), we can then show that the
covariance of $\mu$ is given by:
\begin{equation} \label{eqn_J}
\left(C_{\mu}\right)_{ii^{\prime}} \equiv E\mu_{i}\mu_{i^{\prime}} =
\sum_{j,j^{\prime}}\left(C_{\zeta}\right)_{jj^{\prime}}H_{i}\left(x_{j},y_{j}\right)H_{i^{\prime}}\left(x_{j^{\prime}},y_{j^{\prime}}\right)
+ \left(\sigma_{\nu}\right)_{i}^{2}\delta_{ii^{\prime}}
\end{equation}

The state vector, $\mathbf{\Gamma}$, is also regarded as a
stochastic process; its {\it a priori} probability distribution is
assumed to be described by:
\begin{equation} \label{eqn_K}
P\left(\mathbf{\Gamma}\right) = \prod_{n}P\left(\Gamma_{n}\right)
\end{equation}

\noindent where
\begin{equation} \label{eqn_L}
P\left(\Gamma_{n}\right) = \left\{
\begin{array}
{r@{\qquad}l}
P_{1}   & \mbox{if}~~\Gamma_{n} = 1\\
1-P_{1} & \mbox{if}~~\Gamma_{n} = 0\\
0       & \mbox{otherwise}
\end{array} \right.
\end{equation}

The quantity $P_{1}$ thus represents the {\it a priori} probability
of occupancy of a cell in the state space of flux and position.

\section{Estimation Procedure}

The central operation is to estimate the state vector, $\mathbf{\Gamma}$, given
the observed data.  The details of the estimation process depend on the
type of optimality criterion selected.  For the present algorithm we choose 
a mean square error criterion, for which the optimal estimate is the
{\em a posteriori\/} average value of $\mathbf{\Gamma}$, given by:
\begin{equation}
\rho\left(z_{n}|\mathbf{d}\right) \equiv E\left(\Gamma_{n}|
\mathbf{d}\right) = \sum_\mathbf{\Gamma}\Gamma_{n}P\left(\mathbf{\Gamma}|
\mathbf{d}\right)
\label{eqn_M}
\end{equation}

\noindent where $z_{n}$ is a 3-d vector representing 
the coordinates, $(x_{j},y_{j},f_{k})$, of the $n$th cell in state space, and
the conditional probability, $P(\mathbf{\Gamma}|\mathbf{d})$, is given by
Bayes' rule:
\begin{equation}
P\left({\mathbf\Gamma}|\mathbf{d}\right) = 
P\left(\mathbf{d}|\mathbf{\Gamma}\right) P(\mathbf{\Gamma})/P(\mathbf{d})
\end{equation}

\noindent in which $P({\mathbf\Gamma})$ is given by (\ref{eqn_K}),  
$P(\mathbf{d})$ is a normalization constant, and
\begin{equation}
\ln P\left(\mathbf{d}|\mathbf{\Gamma}\right) = -\frac{1}{2}
(\mathbf{d} - \mathbf{A\Gamma})^{\mbox{\scriptsize{T}}} \mathbf{C}_\mu^{-1}
(\mathbf{d} - \mathbf{A\Gamma}) + {\rm const.}
\end{equation}

We refer to
$\rho\left(z_{n}|\mathbf{d}\right)$ as a density, since
it represents the average local density of occupied cells in the
state space of position and flux. Its estimation
is a generic problem in statistical mechanics; previous applications have
included acoustical imaging \citep{ric87} and target tracking \citep{ric92}.
The solution procedure
involves the solution of a hierarchy of integro-differential
equations which fortunately can be truncated, to a good
approximation, at the first member.  We then obtain:
\begin{equation} \label{eqn_O}
\frac{\partial\rho}{\partial t} + \phi_{1}\rho = 0
\end{equation}

\noindent where $t$ is a dimensionless progress variable
representing the degree of conditioning on the data, and $\phi_{1}$
is the conditioning factor given by:
\begin{equation}
\phi_{1} = -\left(\mathbf{d} -
\mathbf{A}\rho\right)^{\mbox{\scriptsize{T}}}\mathbf{C}_{\mu}^{-1}\mathbf{A}
+ \mathbf{b}/2
\end{equation}

\noindent where $\mathbf{b}$ represents a vector formed from the
diagonal elements of
$\mathbf{A}^{\mbox{\scriptsize{T}}}\mathbf{C}_{\mu}^{-1}\mathbf{A}$.
A discretized version of (\ref{eqn_O}) is formed by regarding the 
process as a series of $N_{\rm c}$ weak conditionings, each of which
corresponds to a measurement noise covariance of $N_{\rm c}\mathbf{C}_\mu$.
The progress variable, $t$, then corresponds to the number of conditionings, 
and the $dt$ in (\ref{eqn_O}) is replaced by 1.
Equation (\ref{eqn_O}) is then to be integrated from $t = 0$ to a
terminal value, $t_{f}=N_{\rm c}$, representing full conditioning on the data.

The initial condition on $\rho$ for the numerical solution of
(\ref{eqn_O}) is $\rho_{(t=0)} = \rho_{0}$, where $\rho_{0}$
is the {\it a priori} density equal to the constant value $P_1$;
its sum over all state space is equal to $N_0$, the {\em a priori\/}
expectation number of planets present.  If we regard the true number of planets,
$N$, as an unknown, then this quantity could, in principle, be estimated
by adjusting $P_1$ for equality between $N_0$ and the {\em a posteriori\/}
number of planets, corresponding to the sum of 
$\rho(z_{n}|\mathbf{d})$ over all state space \citep{ric87}.  
In practice, however, this would be numerically cumbersome.
We have therefore adopted an approximate alternate procedure in which 
the number of planets is estimated with the aid of $\chi_\nu^2$, given by:
\begin{equation} \label{eqn_Q}
\chi_{\nu}^{2} = \frac{1}{M}\sum_i\left(\mathbf{d} -
\mathbf{A}\rho\right)_{i}^{2}/\left(C_{\mu}\right)_{ii}
\end{equation}

\noindent where $M$ is the total number of data samples. In this procedure, we
fix $P_1$ at a very small value, thereby producing an 
{\em a priori\/} bias towards a small number of planets being present.
We then proceed with the numerical integration (during the course of
which, $\chi_\nu^2$ decreases monotonically), terminating it
when $\chi_\nu^2=1$. We thereby arrive at
a solution in which the data are fit by the minimum possible number of planets.

Our corresponding estimate of the source intensity distribution is then:
\begin{equation} \label{eqn_R}
\hat{I}\left(x_{j},y_{j}\right) =
\sum_{k}f_{k}\rho\left(z_{n(j,k)}|\mathbf{d}\right)
\end{equation}

Estimates of the planet fluxes themselves may be
obtained from the integrated value around each peak in this image;
the uncertainties correspond to the standard {\em a posteriori\/} variances
of maximum likelihood estimates.

\section{Tests with Synthetic Data}

\begin{figure}
 \includegraphics[scale=0.64, angle=-90]{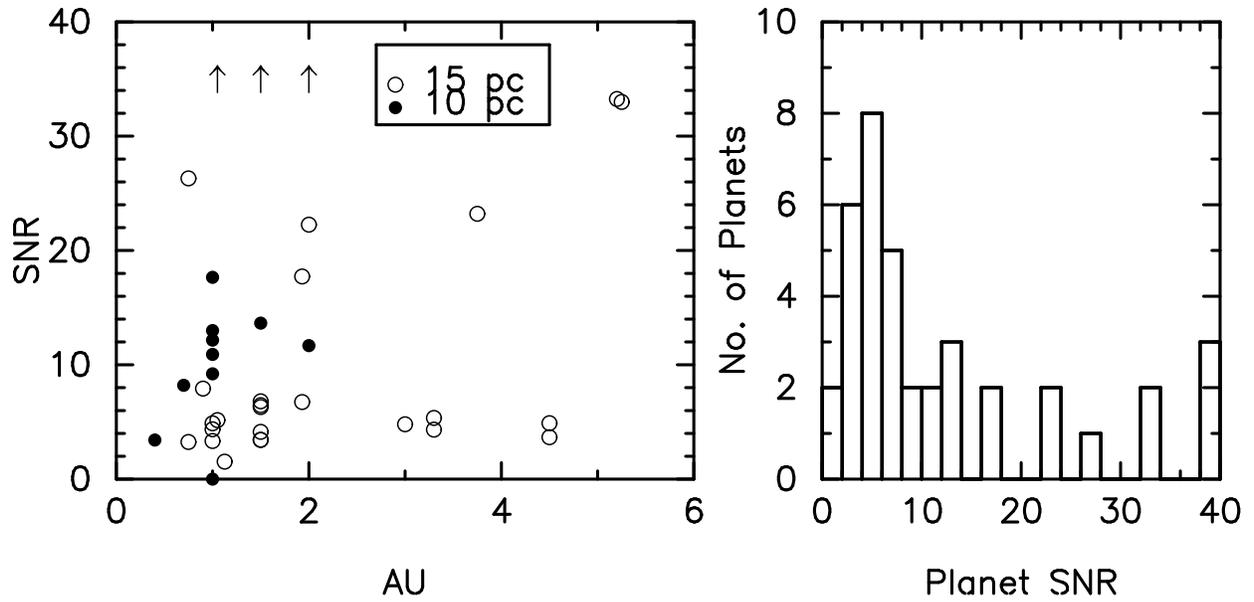}
  \caption{The distribution of planet parameters in the test cases,
with respect to radial distance from the star  and corresponding
ideal SNR for an isolated planet. The distances to the stars are as
indicated.} \label{fig2}
\end{figure}

We have tested the PPA using a set of 15 test cases designed to
replicate realistic observing scenarios for TPF-I in the
wavelength range 7--15 $\mu$m.  Each case involved 0--5 planets at
radial distances of 0.4--5.25 AU from a solar-type star at a
distance of 10 or 15 pc, superposed on a 1 Zodi dust distribution.
The 15 cases involved a total of 37 planets, 24 of which were of 1
Earth flux, and 2 of which were $<1$ Earth flux. The orbital
inclination was $60^\circ$ except for two face-on cases.  All
planets were at Earth temperature (260 K). Radial distances  and
signal-to-noise ratios were distributed as shown in Figure
\ref{fig2}.

The measurement configuration was the ``X array" consisting of
four 4-m detectors at the vertices of a rectangle of width 12 m
(corresponding to the nulling baseline length) and a length of
either 36 m or 72 m, as shown in Figure \ref{fig1}.  
We adopted the 12 m nulling baseline in
lieu of the nominal design value of 20 m to improve the sensitivity at the 
short wavelengths. For longer baselines (20 m in particular), the stellar
leakage becomes larger because of the narrower null, with particularly
severe effects at the shorter wavelengths. 

The measurements consisted of the
simulated sine chop signals through a full rotation of the array,
at wavelengths of 7.44, 8.50, 9.92, 11.90 \& 14.90 $\mu$m, with 
bandwidths ranging from 0.9 $\mu$m at 7.44 $\mu$m to 3.7 $\mu$m at 14.9 $\mu$m,
and a total integration time of 1 day. Superposed on these signals was
Poisson noise due to the various instrumental and astrophysical effects 
discussed by \citet{beichman_velusamy1999}, the most important ones being
stellar leakage (due to incomplete nulling of the
stellar photosphere) and thermal emission from the local- and exo-zodiacal
dust clouds, the latter assumed geometrically symmetrical.

The data were inverted using the PPA, and also by the standard technique
of ``correlation map + CLEAN" for comparison purposes.
The spatial covariance of the exozodi was ignored
(i.e., $C_\zeta$ was set at zero) since
symmetrical sources cancel in the sine chop signal.  Full account was,
however, taken of the Poisson noise contribution of this component.
Figure \ref{fig3} shows the results for five representative cases involving
both array configurations. The results of the PPA and CLEAN algorithms are
shown for the 36 m and 72 m configurations; note that in the case of CLEAN, 
only the 6 strongest source components are shown.  Also shown (Column 4) is the
result of using the PPA with a combination of data from both configurations, 
but with the integration time split equally between the two, i.e., the 
integration time was 0.5 day per configuration.  

Figure \ref{fig4} shows the results for the case involving 5 assumed planets
(Case 4 of the previous figure).  
Also shown for comparison (panel (e) of Figure \ref{fig4}) is the result of 
using noiseless data; it demonstrates that the fidelity of the PPA image is
limited, in part, by the gaps in spatial frequency coverage of
the measurements.

For all 15 test cases, these inversions yielded images from which we extracted
estimates of point source locations, fluxes, and corresponding
uncertainties. We identified the six brightest peaks in each
image, then applied a SNR threshold and counted the detections.
Comparison with the true (assumed) planet positions then led to
the number of true detections and false alarms as a function of
the detection threshold in sigmas. The fluxes of detected sources were
recovered to an accuracy consistent with their theoretical uncertainties.
For the 5-planet case above, with the 36 m array, these uncertainties ranged
from approximately 2\% for the brightest source (8 Earth fluxes) to
20\% for the weakest (1 Earth flux).

Figure \ref{fig5} shows a plot
of true positives vs. false negatives, known as the operating
characteristic curve of the detection system, for both PPA and
CLEAN. Also included in the figure is the theoretical
operating characteristic curve for an isolated source observed with
the 36-m array, assuming pure Gaussian statistics.
It is apparent that the PPA detected significantly more planets 
than CLEAN, particularly in the case of the fainter objects at low $S/N$ 
thresholds. While it is also apparent that the PPA results fall below
the theoretical curve, it should be borne in mind that the latter
did not take into account the interactions between 
sources due to the incomplete spatial frequency coverage.
Figure \ref{fig6} shows the PPA performance as a function of detection 
threshold in sigmas for the 36-m array.

\section{Discussion}

The PPA is theoretically a near-optimal approach, and the present results
support this.  They demonstrate that use 
of the PPA increases the effective sensitivity for planet detection
over standard techniques exemplified by CLEAN.

\begin{figure}
 \includegraphics[scale=0.8, angle=0]{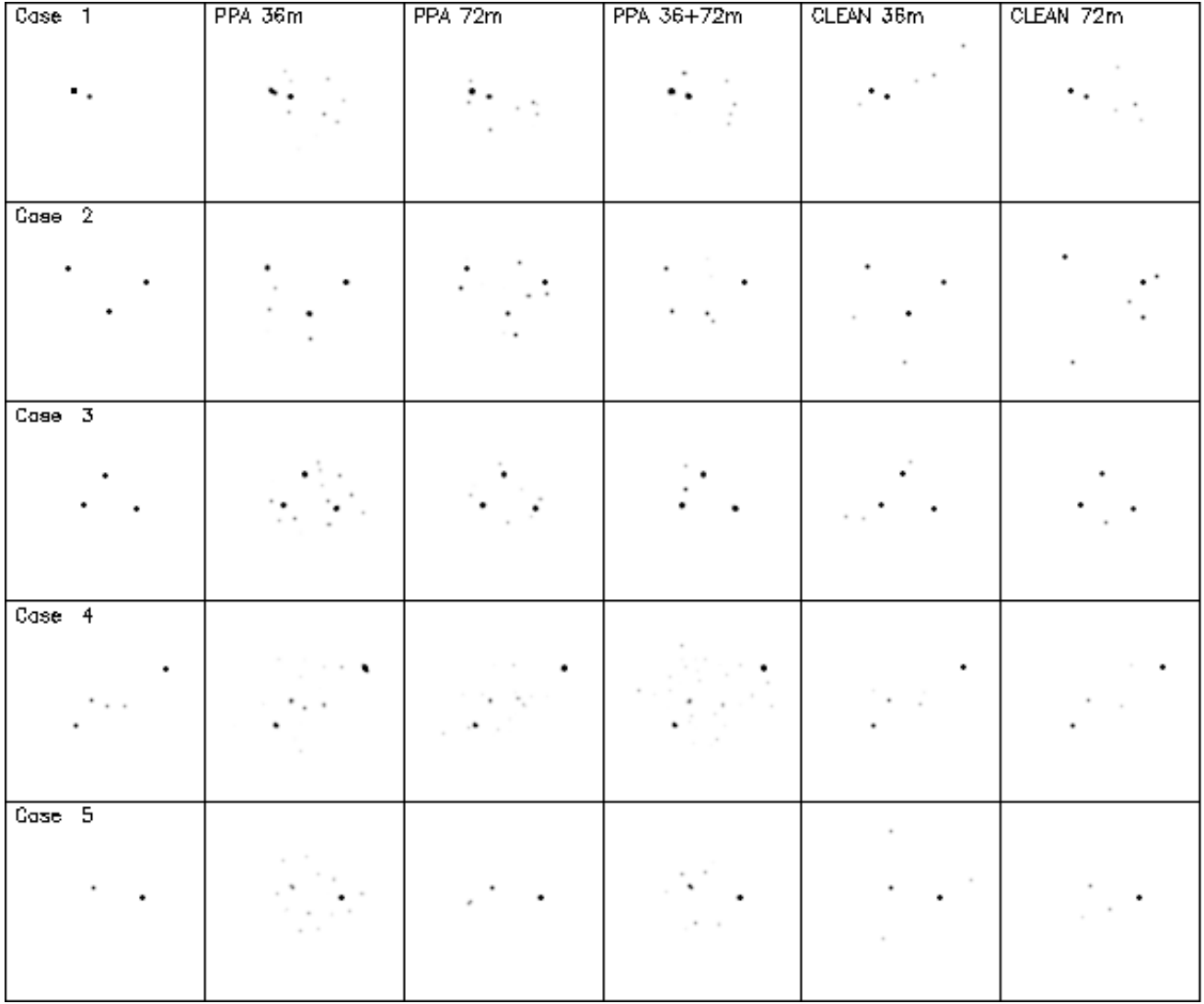}
  \caption{Five representative cases, showing the results for PPA and CLEAN
with two configurations of the interferometer. The true intensity distributions
are shown in the left hand column.  Columns 2 and 3 show the results
of the PPA for the 36 m and 72 m arrays, respectively, for an total 
integration time of 1 day.  Column 4 shows the results of the PPA using
data from both configurations combined, but with the same total integration
time, i.e., a half-day per configuration.  Columns 5 and 6 give
the results obtained using CLEAN for the 36 m and 72 m arrays, showing the
6 strongest source components in each case.} \label{fig3}
\end{figure}

\begin{figure}
 \includegraphics[scale=0.63, angle=-90]{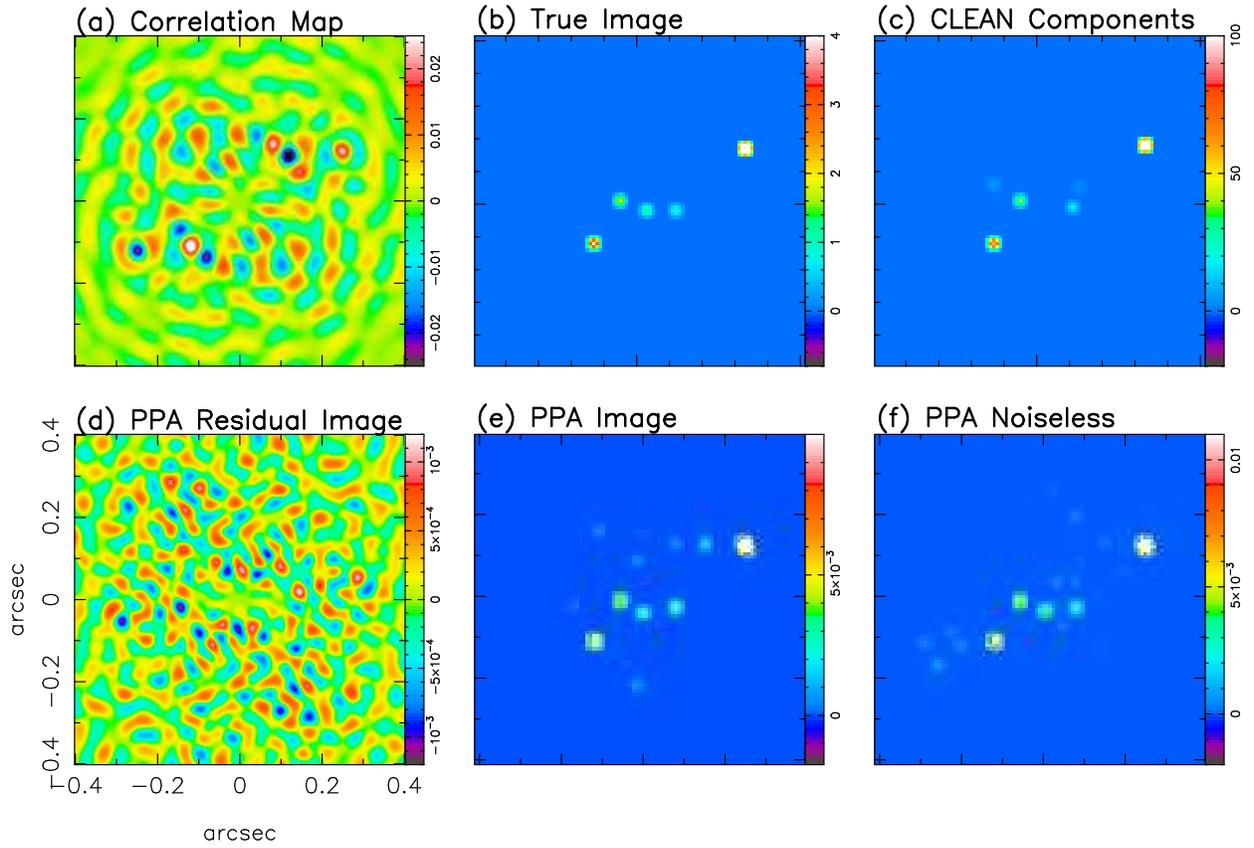}
  \caption{Images obtained by PPA and CLEAN methods for a test case involving 5 planets with the X-36m array. Similar images were obtained for all 15 test cases.}
\label{fig4}
\end{figure}

\begin{figure}
 \includegraphics[scale=0.65, angle=-90]{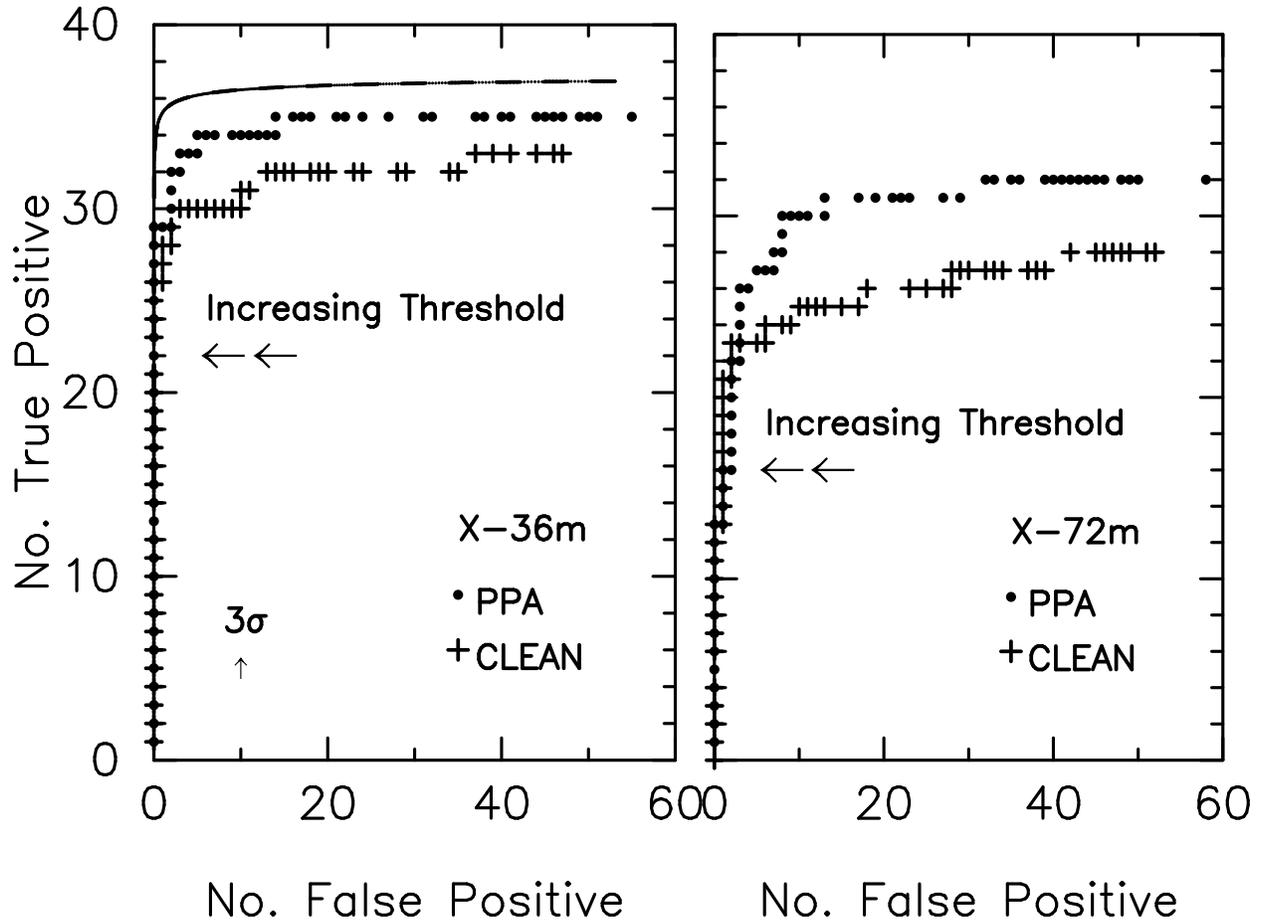}
   \caption{Planet detection statistics, showing the operating characteristic 
curves for PPA and CLEAN. The solid line
on the left hand plot represents the idealized theoretical curve
for an isolated planet.}
\label{fig5}
\end{figure}

\begin{figure}
 \includegraphics[scale=0.8, angle=0]{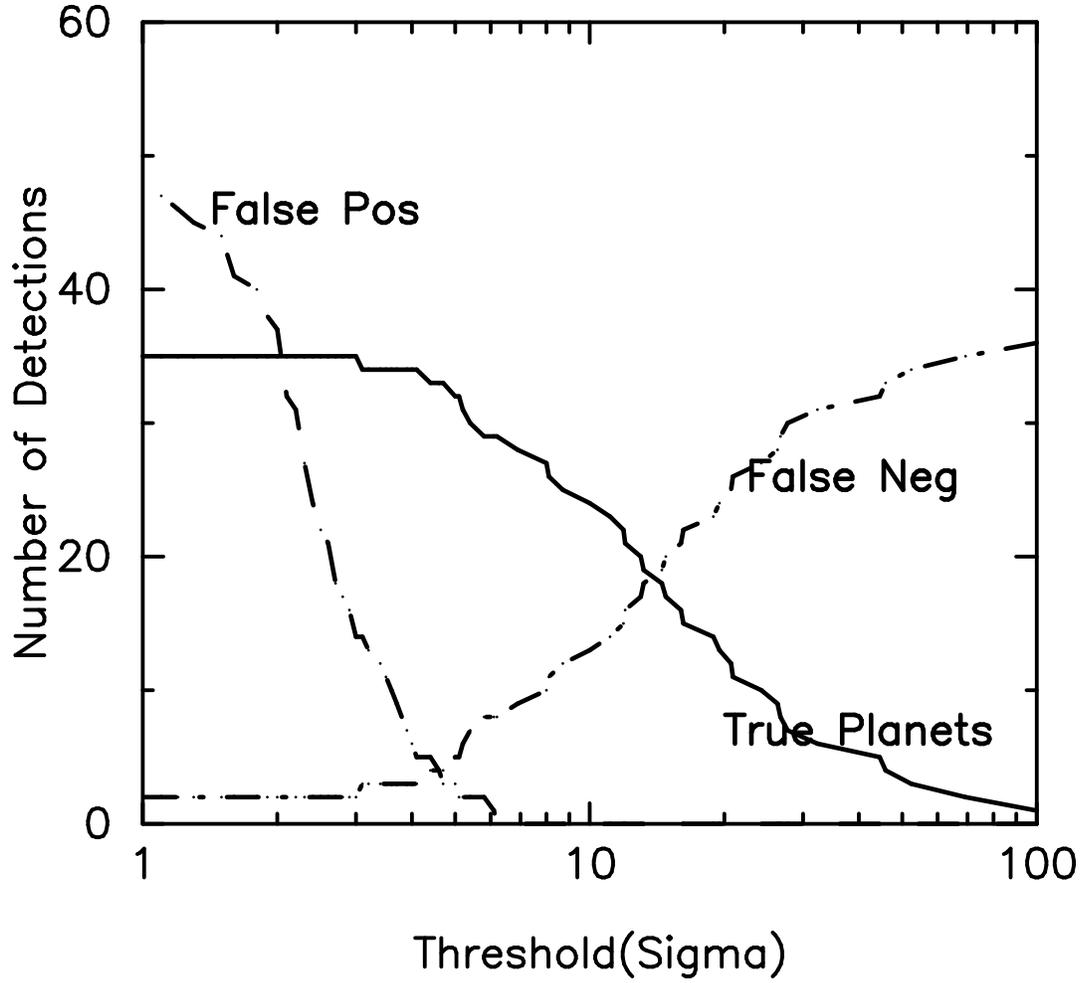}
   \caption{Detection performance of PPA with the 36-m array as a function 
of threshold in sigmas.}
\label{fig6}
\end{figure}

The results also indicate that the 72 m array would be less sensitive than
the 36 m array for this ensemble of planets, presumably due to the smaller
relative range of spatial frequencies involved.  For this reason, it would
be more effective, for a given amount of integration time, to use the 36 m
array exclusively rather than split the time between the 36 m and 72 m
configurations, even though the inclusion of the 72 m array would increase
the overall spatial frequency coverage.  The fact that this added coverage 
did not offset the reduced integration time of the 36 m array is apparent
from the results in Figure \ref{fig3}, and is confirmed by the statistical
results from the complete set of 15 cases.

Future work in the application of the PPA to the planet detection problem
will emphasize the effects of exozodi, including the effects of
spatial inhomogeneities which may masquerade as planets in the interferometer
signals, and their mitigation via appropriate incorporation of
{\em a priori\/} estimates and spatial covariance.  For the latter purpose
we will make use of recent observations of debris disks using the Spitzer
Space Telescope \citep{beichman2005}.

We will also explore some of the possible applications of the PPA beyond the
TPF problem, since, as we have discussed, it is not limited to the detection
of point sources.  It may, in fact, be regarded as a generic technique for
modeling a set of data (which may or may not be image-like) in terms of 
user-defined primitives (i.e., objects) which are parametrizable.  The main 
requirement is that the superposition rule be satisfied, i.e. that 
the measurements represent a linear combination of the contributions from the 
individual objects in the system. The objects themselves need not be defined in
terms of astrophysical entities; they may instead be a set of parametrizable
fractal elements convenient for representing an observed image. In such an
application, the goal of the PPA would be similar to that of adaptive scale 
pixel deconvolution techniques such as the Asp-CLEAN technique of \cite{bha04} 
and the ``pixon" technique \citep{dix96}. Unlike the latter, however, the PPA 
is not restricted to cases involving PSFs of finite support, and it has some 
potential advantages over the former. One of these is that it may provide a 
more efficient procedure for spatially-complex images, since a consequence of 
the PPA's use of an occupation-number representation is that the computational 
burden does not increase significantly with the number of source components in 
the image.  The second potential advantage of the PPA is that it is fully 
capable of superresolution, which in practical terms means that it can separate
adjacent point sources whose separation is less than the PSF width.  It is not 
clear that Asp-CLEAN would have this capability given the fact that the standard
CLEAN algorithm does not.  We will investigate these and other issues in
order to better characterize PPA's niche in astronomical problems.

%\newpage

\acknowledgments

We thank the referee for helpful comments.
This work was performed by the Jet Propulsion Laboratory, 
California Institute of Technology, under contract with the 
National Aeronautics and Space Administration.

\clearpage

\clearpage

\clearpage

\clearpage

\clearpage

\clearpage

\end{document}